\documentstyle[11pt,paspconf,epsf]{article}

\begin{document}

\title{Magnetically-Driven Winds from Protostellar Disks}

\author{Mark Wardle}
\affil{Research Centre for Theoretical Astrophysics, University of Sydney}

\begin{abstract}
Angular momentum transport in protostellar disks
can be achieved by the action of a large scale magnetic field that runs
vertically through the disk.  The magnetic field centrifugally drives
material from the disk surfaces into a wind, initiating a bipolar outflow.
One apparent difficulty for this model is that the conductivity of the disk
is extremely low in the inner 0.1--10 AU of the disk, where grains are the
dominant charge carriers.  Near the midplane, charged grains are unable to
drift through the neutral gas and there is negligible coupling between the
magnetic field and the disk material.

However, the removal of angular momentum and acceleration of a wind by a
magnetic field can still take place in the surface layers of the disk
where the gas conductivity increases dramatically.  Solutions to the 
multifluid MHD equations for the vertical structure of a disk at a
particular radius are presented.  Most of the disk material sits in
hydrostatic equilibrium and does not interact with the magnetic field
running vertically through it.  Near the disk surfaces, the coupling
between the magnetic field and disk material is sufficient to initiate an
outflow from the disk surface.

\end{abstract}



\section{Introduction}

Disk-wind models for protostellar outflows are based on the
self-similar solutions of Blandford and Payne (1982).  Open magnetic
field lines threading the disk vertically are carried around the
central object, and material is centrifugally flung out along the
field lines to form an outflow.  The Blandford and Payne solutions and
the numerical simulations (Ustyugova et al.  1995;  Meier, Payne and
Lind 1996; Ouyed, Pudritz and Stone 1996), do not explicitly include
the disk, instead assuming a steady injection of disk material at the
base of the wind, and that the footpoints of the field lines are
frozen into a Keplerian disk.

The microphysics determining the disk's conductivity is an integral
part of the problem of accelerating outflows.  Flux-freezing must
break down within the disk, otherwise the magnetic field would be
continuously advected towards the central object while being amplified
by differential rotation.   The finite conductivity of the disk is
also intimately related to the loading of some of the material onto
field lines at the disk surface and its injection into the wind
(Wardle and K\"onigl 1993).  Perhaps most importantly, the disk
conductivity places constraints on the ratio of the radial and
azimuthal components of the magnetic field at the disk surface.  For
example, in the absence of radial field line drift the electric field
within the disk is predominantly radial.  The disk conductivity, a
tensor, determines the ratio of the radial and toroidal current
flowing within the disk in response to the electric field.  The
current, in turn, determines the relative magnitude of the toroidal
and radial field components at the disk surfaces.   Thus the disk
conductivity effectively imposes an additional boundary condition that
has yet to be incorporated into the wind models.

\begin{figure}
\plotone{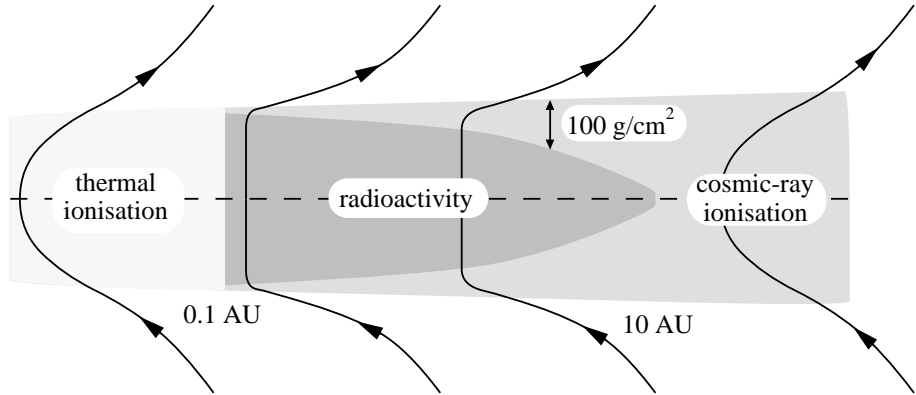}
\caption{A sketch of the ionisation structure and magnetic field
configuration expected for a protostellar disk.  Cosmic rays can
penetrate the first 100 g/cm$^2$ of the disk, so cannot maintain
significant ionisation deep into the disk inside about 10 AU.  Thermal
ionisation of metals becomes significant within 0.1 AU where the disk
temperature exceeds 1000 K.  The magnetic field does not interact with
the disk material in the region where radioactivity is the dominant
form of ionisation because the conductivity there is negligible.}
\label{fig-1}
\end{figure}

\section{The conductivity and structure of protostellar disks}

Wardle and K\"onigl (1993; herafter WK) examined the effect of the
breakdown of flux-freezing in the outer regions of protostellar disks,
beyond a few tens of AU from the central star.  At these radii the
dominant charged species are ions and electrons, and (as suggested by
K\"onigl 1989) the conductivity is low enough to permit significant
ambipolar diffusion, but not so low that the coupling between the
field and gas is negligible.   WK constructed solutions for the
vertical structure of the disk that could then be matched
approximately onto self-similar wind solutions.  Li (1996) has recently
constructed analogous self-similar disk-wind solutions.

Several factors cause the conductivity to plummet within 10 AU of the
protostar as the gas density becomes successively higher.  There are
more neutral targets for drifting charged species to collide with;
dust grains, which have a large neutral collision cross-section,
become the dominant charged species (e.g. Nishi, Nakano and Umebayashi
1991); and the gas is self-shielding from cosmic rays once the surface
density of the disk is larger than about 200 g cm$^{-2}$, reducing the
number of charged particles per unit volume.  All of these effects
reduce the current that can flow in the presence of an applied
electric field.  The conductivity  partially recovers within about 0.1
AU of the protostar where the disk temperature becomes sufficient for
to ionise potassium (Hayashi 1981; see also Li 1996).

Although the conductivity at the disk midplane is far too low to
couple the field and disk material between 0.1 and 10 AU, cosmic rays
can penetrate the disk to a depth of 100 g/cm$^2$, and there may be
sufficient coupling between the field and the surface layers of the
disk for magnetic effects to play an important role (Hayashi 1981;
Gammie 1996). A plausible magnetic field structure is sketched in
Figure 1.
The magnetic field lines are straight in the ``dead'' region of very low
conductivity, which cannot support a significant current.  The surface
layers, however, can support an accretion flow and accelerate an disk
wind in much the same way as the original Wardle and K\"onigl
solutions.

New solutions have been obtained using the WK equations with some
modifications: the charged species are dust grains
rather than ions and electrons, the number density of
charged particles scales as the square root of the gas density, rather
than being constant, and the cosmic-ray ionisation rate decreases
exponentially with depth into the disk, with a characteristic
attenuation column of 96 g cm$^{-2}$.

Figure 2 compares a typical WK solution
(left panel) with a solution appropriate to conditions at 1 AU (right panel).
\begin{figure}
\plotfiddle{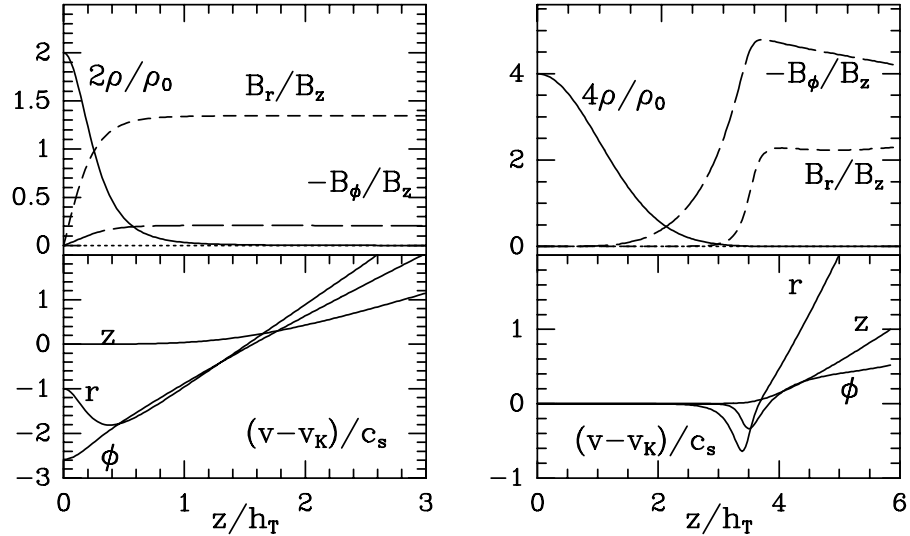}{7cm}{0}{60}{60}{-194}{-80}
\caption{Vertical structure of a protostellar disk outside of 10 AU
(left) and at 1 AU (right).  The top panel of each figure plots the
run of density and of the magnetic field components with height above the
midplane in units of the isothermal scale height.  The velocity components
(relative to Keplerian and normalised by the sound speed) are
plotted in the lower panels.} \label{fig-2}
\end{figure}
As expected, the new solutions show that the bulk of the disk material
is inactive.  Cosmic rays can maintain a reasonable ionisation
fraction in the surface layers, where the solution resembles the WK
solutions; the magnetic field bending away from the rotation axis and
centrifugally driving an outflow from the disk surface through a sonic
point.

\section{Discussion}

There are some important differences between the two solutions.
Firstly, in the WK solutions, the disk is compressed by magnetic
stresses, so is appreciably smaller than the isothermal scale height.
This is not the case in the new solution, where the field pressure is
much smaller than the thermal pressure at the midplane and most of the
disk is unaffected by magnetic stresses.  Secondly, Wardle and
K\"onigl showed that in the ambipolar diffusion dominated case, the
toroidal field dominated the poloidal field at the disk surface.  In
the present case, the {\it opposite} is true.

The solution for the vertical structure at 1AU is not strictly correct
since by the time the material passes through the sonic point, the gas
density is of order $10^{10}$ cm$^{-3}$, comparable to the midplane
density at 30 AU, where ions and electrons are again important.
Work towards solutions incorporating multiple charged species 
(i.e. ions, electrons, and charged grains) is underway.

\acknowledgments This work forms part of a collaborative effort with
Arieh K\"onigl.


\begin{references}
\reference Blandford, R. D., \& Payne, D. G. 1982, \mnras, 199, 883
\reference Gammie, C. F., 1996, \apj, in press
\reference Hayashi, C. 1981,  Prog. Theor. Phys. Suppl. 70, 35
\reference K\"onigl, A. 1989, \apj, 342, 208
\reference Li, Z.-Y. 1996, \apj, 465, 855
\reference Meier, D. L., Payne, D. G., \& Lind, K. R. 1996, in
   Extragalactic Radio Sources, in press
\reference Nishi, R., Nakano, T., \& Umebayashi, T. 1991, \apj, 368, 181
\reference Ouyed, R., Pudritz, R. E., \& Stone, J. M. 1996, Nature, submitted
\reference Ustyugova, G. V., Koldoba, A. V., Romanova, M. M., 
   Chechetkin, V. M., \& Lovelace, R. V. E. 1995, \apjlett, 439, L39
\reference Wardle, M., \& K\"onigl, A. 1993, \apj , 410, 218

\end{references}
\end{document}